# Geometric Journal Impact Factors Correcting for Individual Highly Cited Articles[1]

Mike Thelwall
Statistical Cybermetrics Research Group, School of Mathematics and Computer Science, University of Wolverhampton, Wulfruna Street, Wolverhampton WV1 1LY, UK.
Tel. +44 1902 321470. Fax +44 1902 321478, Email: m.thelwall@wlv.ac.uk
Ruth Fairclough
Statistical Cybermetrics Research Group, School of Mathematics and Computer Science, University of Wolverhampton, Wulfruna Street, Wolverhampton WV1 1LY, UK.
Tel. +44 1902 321000. Fax +44 1902 321478, Email: r.fairclough@wlv.ac.uk

Journal Impact Factors (JIFs) are widely used and promoted but have important limitations. In particular, JIFs can be unduly influenced by individual highly cited articles and hence are inherently unstable. A logical way to reduce the impact of individual high citation counts is to use the geometric mean rather than the standard mean in JIF calculations. Based upon journal rankings 2004-14 in 50 sub-categories within 5 broad categories, this study shows that journal rankings based on JIF variants tend to be more stable over time if the geometric mean is used rather than the standard mean. The same is true for JIF variants using Mendeley reader counts instead of citation counts. Thus, although the difference is not large, the geometric mean is recommended instead of the arithmetic mean for future JIF calculations. In addition, Mendeley readership-based JIF variants are as stable as those using Scopus citations, confirming the value of Mendeley readership as an academic impact indicator.

## Introduction

JIFs are probably the most used and misused scientometric indicators. The Thomson Reuters JIF for the year X is the number of citations from Web of Science (WoS)-indexed documents published in year X to journal articles published in years X-1 and X-2 divided by the number of articles published in the journal in years X-1 and X-2 (Thomson Reuters, 2014). It is thus an indicator of the average citation rate of recent articles in a journal. Other JIFs use different citation databases (e.g., Scopus) and different timeframes (e.g., 5 years) and different concepts of citable items. JIFs seem to be displayed on the home pages of almost all journals that have them and are an important part of the annual reports sent by some publishers to editorial boards. Some countries also use JIFs to help their funding model, for example by using them to aid peer review in the ranking of journals. Moreover, researchers in Spain (Brown, 2007) and elsewhere (Al-Awqati, 2007) can be directly rewarded for publishing in high JIF journals. Less formally, it seems likely that a substantial fraction of researchers use JIFs to help decide where to publish and a substantial minority of hiring and promotion decisions are influenced by the JIFs of candidates' publications. JIFs may also be important for generating citations because the journal in which an article is published appears to be an important indicator of its value to the scientific community and therefore plays an important role in publicising the article. As evidence of this, one study found that 4,532 published articles tended to attract twice as many citations as similar articles (with same first author, title and references) published in journals with lower JIFs (Larivière & Gingras, 2010).

   Thomson Reuters clearly signals the limitations of the JIF on its website, such as "The impact factor can be used to provide a gross approximation of the prestige of journals in which individuals have been published. This is best done in conjunction with other considerations such as peer review, productivity, and subject specialty citation rates" (Thomson Reuters, 2014) and more bluntly, "Thomson Reuters does not depend on the impact factor alone in assessing the usefulness of a journal, and neither should anyone else" (Thomson Reuters, 2014). Nevertheless, this advice seems to be widely ignored by busy scientists and research managers (Adler & Harzing, 2009) and, perhaps as a result of this, the JIF has been frequently criticised (Archambault, & Larivière, 2009; Seglen, 1997). This has led to the formation of the San Francisco Declaration on Research Assessment (DORA), which argues that the Thomson Reuters JIF should not be used because of four main limitations: the skewed nature of citations; field differences in citation counts and the mixing of different publication types; gaming by journals; and a lack of transparency (DORA, 2012). Two additional concerns are the lack of confidence intervals and the use of decimal places indicating a spurious accuracy level (Vanclay, 2012) as well as biases in the citation indexes from which they are derived and uncritical uses of them that do not take into account field differences in citation norms (Seglen, 1997; Archambault, & Larivière, 2009). In addition, JIF calculations count all citations as equal, but citations from more important articles should perhaps carry a higher weight and this can be achieved with network-based alternatives to the JIF, such as those based upon eigenvalues of network matrices (Bergstrom, West, & Wiseman, 2008).

---

[1] Thelwall, M. & Fairclough, R. (2015). Geometric journal impact factors correcting for individual highly cited articles, Journal of Informetrics, 9(2), 263–272.

In partial support of the validity of JIFs, peer review ratings have been shown to correlate with JIFs in nine fields in an Italian study (mathematics and computer sciences, chemistry, earth sciences, biology, medical, agricultural sciences and veterinary medicine, civil engineering and architecture, industrial and information engineering, economics and statistics). The correlation was not statistically significant for physics, however, and the social sciences and humanities were not tested, except for economics (Franceschet & Costantini, 2011). A smaller-scale regional study has found little relationship, however (Haddow & Genoni, 2010), and another found expert journal rankings to be biased towards the researcher' current interests (Serenko & Dohan, 2011).

Perhaps the two most fundamental criticisms are that JIFs are not credible guides to the impact of individual articles and that they can be greatly influenced by individual highly cited articles (e.g., one article has temporarily increased the publishing journal JIF from 2 to 50 Foo, 2013). The first argument mainly concerns the use of JIFs in research evaluation, claiming that it is better to consider the citation counts of individual articles when evaluating, say, a researcher, than to consider the JIFs of the publishing journals. This is because the former is a more direct impact indicator and the skewed nature of citation distributions means that JIFs are not a good approximation of the citation counts for many of the articles in a journal (Lozano, Larivière, & Gingras, 2012; Seglen, 1992, 1994). Nevertheless, this argument is not strong in itself because it is also possible that the quality of the journal in which an article is published is a better guide to the quality of an article than is its citation count. This is based upon the assumption that scientific quality is not the same as citation impact, which seems clear because there are many reasons when an article may be highly cited that do not relate to its scientific quality as judged by its peers (MacRoberts & MacRoberts, 1989). This distinction is perhaps clearest in high quality general journals that may publish articles from fields with both high and low citation norms. It seems unlikely that, for example, uncited *Nature* papers would be considered to be poor. The importance of the journal is operationalised in journal rankings or journal quality categories constructed by peers, such as those previously used in the Excellence in Research Australia (ERA) funding exercise (UNSW, 2015). One conference ranking system requires "detailed information [] about paper acceptance rates, the Local and Program Committee Chairs, and other indicators of quality" (CORE, 2014, p.2) as well as citation information in order to decide how to rank a computer science conference, suggesting faith in a combination of the peer review process and citation counts. Ranked journal lists are also commonly produced in disciplines, reflecting a belief that the choice of journal for an article can be an important indicator of its value (e.g., ABS, 2010; Kalaitzidakis, Mamuneas, & Stengos, 2003, 2011; Nisonger & Davis, 2005; Rainer & Miller, 2005; Steward & Lewis, 2010). A belief that the journal in which an article is published in is an important (and convenient) indicator of its value is presumably the main reason why JIFs continue to be widely used, despite their acknowledged limitations at indicating the quality of a journal.

The second fundamental criticism of JIFs is that they can be greatly influenced by small numbers of highly cited articles (Seglen, 1992). For example, one highly cited article in a small journal may double or triple the journal Thomson Reuters JIF during the two years in which the article is included in the JIF calculation, without any change in the average impact of the remaining articles in the journal. Small numbers of highly cited articles are to be expected because the distribution of citation counts to individual articles is highly skewed (Seglen, 1994), following a hooked power law or lognormal distribution (Thelwall & Wilson, 2014). Evidence about the influence of highly cited articles on JIFs has led to particularly strong condemnation (e.g., Baum, 2011).

A logical way to reduce the influence of individual highly cited articles on JIFs is to average using the geometric mean rather than the arithmetic mean (Zitt, 2012). The geometric mean averages the logs of a set of numbers rather than the numbers themselves, thus reducing the influence of very high values. The use of logs in response to skewed citation data has been previously proposed for individual articles and field normalisation benchmarks (Lundberg, 2007) as well as for whole journals (Stringer, Sales-Pardo, & Amaral, 2008). The latter study analysed the 2,267 journals that had published at least 50 articles in WoS per year for at least 15 years, showing that the distribution of the log of the citation counts of articles in a journal tends to stabilise after 10 years and proposes the mean of this distribution as the key factor for ranking journals, effectively generating a very stable 10-year logarithmic JIF (Stringer, Sales-Pardo, & Amaral, 2008) but which cannot be sensitive to underlying changes in the importance of a journal. Previous scientometric studies have also used the geometric mean for a variant of the h-index (Moussa & Touzani, 2010), for library data (Robinson, & Smyth, 2008), for questionnaires (Donohue & Fox, 2000) and as part of a modelling approach to the standard JIF (Greenwood, 2007). This article assesses whether the geometric JIF, or gJIF, is an improvement on the standard JIF with a comparison across a range of different fields. The basis for the comparison is purely the stability of the results for individual journals over time.

It is also possible and practical to calculate journal impact indicators from alternative types of data, such as from parts of the web (Thelwall, 2012), from usage data (Shepherd, 2007) or by counting mentions in the social web (Priem, Taraborelli, Groth, & Neylon, 2010). Although social web data has been widely used to calculate article-level metrics (Adie & Roe, 2013; Shema, Bar-Ilan, & Thelwall, 2014; Thelwall, Haustein, Larivière, & Sugimoto, 2013; Zahedi, Costas, & Wouters, 2014), journal-level metrics are also possible. For example, a variant of the JIF, called journal usage intensity, has been calculated with readership data from the social reference sharing site Mendeley for articles in 45 physics journals 2004-2008 (Haustein & Siebenlist, 2011). The results had mostly low or moderate correlations with a range of other indicators. Against the traditional JIF, the rank correlation was only 0.136. Mendeley is currently the most promising web-based alternative article-level impact metric because readership counts have a high correlation with citation counts (Li, Thelwall, & Giustini, 2012), and because people who register as readers of articles in Mendeley seem to be genuine readers (Mohammadi, Thelwall, & Kousha, in press) although they are predominantly junior academics and doctoral students (Mohammadi, Thelwall,

Haustein, & Larivière, in press). An additional advantage for JIFs is that Mendeley readership accumulates more rapidly than do citations (Maflahi & Thelwall, in press; Thelwall & Wilson, in press), and so Mendeley JIFs could be more timely than citation-based JIFs. As a second research direction, then, this article assesses the stability over time of JIFs calculated from Mendeley data.

## Research questions

The objective of this paper is to assess whether journal rankings based on JIFs would be more stable with the geometric mean instead of the standard mean. Whilst this is clear at an abstract mathematical level, it is important to test with real data. In terms of stability, it seems reasonable to use ranking changes over time as the key measure since it is unlikely that the characteristics of journals often change dramatically over time. A possible exception is that changes in editor, whole editorial board or direction may occasionally occur for journals but this does not seem likely to be a frequent occurrence. Since disciplinary differences are important in research, it is useful to assess whether the answer is likely to vary between subjects. The following research questions drive the study.

1. Do journal rankings based upon citation counts for journal articles more stable over time if the geometric mean is used rather than the arithmetic mean?
2. Do journal rankings based upon Mendeley reader counts for journal articles more stable over time if the geometric mean is used rather than the arithmetic mean?
3. Are journal rankings based upon Mendeley reader counts for journal articles less stable over time than journal rankings based upon citation counts?
4. Do the answers to the above vary by discipline?

## Methods

To answer the above questions, a set of data for citation counts to journal articles is needed for a variety of disciplines and years. Scopus was chosen for the citation count data. Although the Thomson Reuters JIFs are better known than the Scopus equivalents, the wider coverage of Scopus (Bartol, Budimir, Dekleva-Smrekar, Pusnik, & Juznic, 2014; Moed & Visser, 2008) is an advantage for testing purposes is because of the higher citation counts that it is likely to generate (Haddow & Genoni, 2010; Torres-Salinas, Lopez-Cózar, & Jiménez-Contreras, 2009). The following Scopus categories were chosen to represent a varied range of subjects: Agricultural and Biological Sciences; Business, Management and Accounting; Decision Sciences; Pharmacology, Toxicology and Pharmaceutics; and Social Sciences. The time period was chosen to be 2004-2014 and the citation counts were downloaded from 18 November 2014 to 16 December 2014, with a maximum of 10,000 articles per year and sub-category studied. Mendeley reader counts were extracted from the Mendeley API from 20 November 2014 to 18 December 2014. Articles were identified in Mendeley with a Digital Object Identifier (DOI) query (Zahedi, Haustein, & Bowman, 2014) in the free Webometric Analyst software, when a DOI was available in Scopus, as well as a search by title, year and first author last name, automatically checking the results returned by Mendeley for accuracy (for more details see the methods section of: Thelwall & Wilson, in press). The reader counts of all matching articles were totalled in cases where there was more than one match (i.e., where different readers had recorded the same article separately).

For each journal in the dataset and for each year, two JIFs were calculated. The arithmetic JIF (called aJIF henceforth) is the arithmetic mean of the citation counts for all the documents of type article in the journal and year. Similarly, the geometric JIF (called gJIF henceforth) was calculated in the same way except for the use of the geometric mean instead of the arithmetic mean. For the geometric mean, 1 was added to the citation counts for each article, then the log of the result taken. The antilog was then taken of the mean of the logs for all articles in the journal and 1 was subtracted from the result, as shown below for a set of $n$ articles, with the $i$th article having $c_i$ citations.

$$gJIF = \exp\left(\frac{\sum_{i=1}^{n} \log(1 + c_i)}{n}\right) - 1$$

The procedure of adding 1 before taking the logs avoids the problem of taking the log of zero in the geometric calculation and is a standard variant. Note that the gJIF and aJIF differ from the Thomson Reuters JIF in that they use citation counts for articles from only one year (rather than citation counts for articles from two years) and they use citations from all articles (rather than from articles in just one year). Both differences (and the swap from WoS to Scopus) should make the calculations more unstable due to the focus on citations to a smaller set of articles from a larger set of articles, which increases the scope for individual high citation scores to change the results. Since the results could be unduly affected by journals with small numbers of articles, the calculations were repeated after excluding from the rankings any journal that had published less than 10 articles in a year.

For each subcategory of each major category above, and for each year, two ranked lists of the journals were created, one using the aJIF and the other using the gJIF. To measure the stability of the ranking in each case, a Spearman correlation was calculated for the rankings of consecutive years (e.g., the journal ranking for 2005 was correlated with the journal ranking from 2004), ignoring journals with fewer than ten articles in the subject category for one or both years. These correlations were calculated for each consecutive pair of years and then the results averaged.

To give an idea of the magnitude of the numbers involved, all the data sets together contained a total of 22,470,090 citations and 24,596,045 Mendeley readers. The mean number of readers per article was 9.3 and the mean number of citations per article was 7.6. Although there were more readers than citations overall, this varied considerably between disciplines, from Drug Discovery (12.5 citations and 4.7 readers per article) to Business, Management and Accounting (misc.) (3.8 citations and 8.4 readers per article).

## Results

Tables 1-5 show the average correlations between years. Although for the citation data the geometric JIF tends to be better than the arithmetic JIF in most or all sub-categories of each broad category, the differences are not large and the rankings are quite stable over time in both cases. The geometric JIF also tends to be slightly better than the arithmetic JIF for the data based on Mendeley readership. Finally, the Mendeley readership-based gJIF is about as stable as the Scopus citations-based gJIF, although in some broad areas one seems to be significantly more stable than the other.

Table 1. Correlations over time between journal rankings based upon arithmetic and geometric JIFs for the Scopus category of Agricultural and Biological Sciences. In each case the correlation is the average Spearman correlation between all ten pairs of consecutive years from 2004 to 2014.

| Sub-category* | Scopus citations | | | | Mendeley readers | | | |
|---|---|---|---|---|---|---|---|---|
| | aJIF | gJIF | aJIF 10+ | gJIF 10+ | aJIF | gJIF | aJIF 10+ | gJIF 10+ |
| Agricultural and Biological Sciences (misc.) | 0.911 | <u>**0.919**</u> | 0.940 | **0.947** | 0.887 | **0.900** | 0.911 | **0.947** |
| Agronomy and Crop Science | 0.891 | <u>**0.891**</u> | 0.925 | **0.931** | 0.851 | **0.865** | 0.920 | **0.930** |
| Animal Science and Zoology | **0.828** | 0.827 | 0.880 | **0.888** | 0.827 | <u>**0.840**</u> | 0.912 | **0.935** |
| Aquatic Science | 0.868 | **0.870** | 0.893 | **0.897** | 0.871 | <u>**0.874**</u> | 0.897 | **0.906** |
| Ecology, Evolution, Behavior and Systematics | 0.832 | **0.839** | 0.882 | **0.890** | 0.871 | <u>**0.884**</u> | 0.920 | **0.937** |
| Food Science | 0.865 | <u>**0.872**</u> | 0.919 | **0.922** | 0.822 | **0.833** | 0.901 | **0.909** |
| Forestry | **0.911** | <u>0.910</u> | 0.923 | **0.925** | 0.850 | **0.861** | 0.910 | **0.923** |
| Horticulture | 0.868 | <u>**0.879**</u> | 0.892 | **0.903** | 0.846 | **0.856** | 0.927 | **0.943** |
| Insect Science | 0.897 | <u>**0.911**</u> | 0.902 | **0.917** | 0.882 | **0.884** | 0.919 | **0.928** |
| Plant Science | 0.875 | <u>**0.876**</u> | 0.914 | **0.920** | 0.841 | **0.844** | 0.922 | **0.933** |
| Soil Science | 0.927 | <u>**0.931**</u> | 0.936 | **0.936** | 0.868 | **0.876** | 0.943 | **0.951** |

*The highest out of the aJIF and gJIF is in bold, and the highest out of the Mendeley and Scopus gJIF is underlined.
+ The 10+ JIF variants exclude journals publishing less than 10 articles in a given year.

Table 2. Correlations over time between journal rankings based upon arithmetic and geometric JIFs for the Scopus category of Business, Management and Accounting. In each case the correlation is the average Spearman correlation between all ten pairs of consecutive years from 2004 to 2014.

| Sub-category* | Scopus citations | | | | Mendeley readers | | | |
|---|---|---|---|---|---|---|---|---|
| | aJIF | gJIF | aJIF 10+ | gJIF 10+ | aJIF | gJIF | aJIF 10+ | gJIF 10+ |
| Business, Management and Accounting (misc.) | 0.799 | <u>**0.815**</u> | 0.820 | **0.825** | 0.804 | **0.814** | 0.849 | **0.881** |
| Accounting | **0.847** | 0.839 | 0.863 | **0.859** | 0.886 | <u>**0.905**</u> | 0.928 | **0.945** |
| Business and International Management | 0.886 | **0.896** | 0.901 | **0.915** | 0.883 | <u>**0.905**</u> | 0.918 | **0.943** |
| Management Information Systems | 0.858 | **0.863** | 0.872 | **0.875** | 0.891 | <u>**0.930**</u> | 0.907 | **0.944** |
| Management of Technology and Innovation | 0.864 | **0.882** | 0.880 | **0.904** | 0.902 | <u>**0.920**</u> | 0.923 | **0.944** |
| Marketing | 0.880 | **0.886** | 0.880 | **0.895** | 0.894 | <u>**0.920**</u> | 0.887 | **0.926** |
| Organizational Behavior and Human Resource Management | 0.870 | **0.872** | **0.883** | 0.882 | 0.906 | <u>**0.921**</u> | 0.899 | **0.919** |
| Strategy and Management | 0.868 | **0.876** | 0.887 | **0.901** | 0.894 | <u>**0.909**</u> | 0.927 | **0.944** |
| Tourism, Leisure and Hospitality Management | 0.779 | **0.781** | 0.819 | **0.826** | 0.808 | <u>**0.832**</u> | 0.885 | **0.895** |
| Industrial Relations | **0.881** | <u>0.880</u> | 0.892 | 0.873 | 0.724 | **0.735** | 0.862 | **0.898** |

*The highest out of the aJIF and gJIF is in bold, and the highest out of the Mendeley and Scopus gJIF is underlined.
+ The 10+ JIF variants exclude journals publishing less than 10 articles in a given year.

Table 3. Correlations over time between journal rankings based upon arithmetic and geometric JIFs for the Scopus category of Decision Sciences. In each case the correlation is the average Spearman correlation between all ten pairs of consecutive years from 2004 to 2014.

|  | Scopus citations | | | | Mendeley readers | | | |
|---|---|---|---|---|---|---|---|---|
| Sub-category* | aJIF | gJIF | aJIF 10+ | gJIF 10+ | aJIF | gJIF | aJIF 10+ | gJIF 10+ |
| Decision Sciences (misc.) | - | - | - | - | - | - | - | - |
| Information Systems and Management | 0.852 | **0.872** | 0.869 | **0.883** | 0.907 | **0.925** | 0.936 | **0.953** |
| Management Science and Operations Research | 0.835 | **0.859** | 0.848 | **0.882** | 0.890 | **0.904** | 0.929 | **0.941** |
| Statistics, Probability and Uncertainty | 0.817 | **0.832** | 0.848 | **0.863** | 0.797 | **0.869** | 0.895 | **0.931** |

*The highest out of the aJIF and gJIF is in bold, and the highest out of the Mendeley and Scopus gJIF is underlined.
+ The 10+ JIF variants exclude journals publishing less than 10 articles in a given year.
-The first category had no journals in the early years.

Table 4. Correlations over time between journal rankings based upon arithmetic and geometric JIFs for the Scopus category of Pharmacology, Toxicology and Pharmaceutics. In each case the correlation is the average Spearman correlation between all ten pairs of consecutive years from 2004 to 2014.

|  | Scopus citations | | | | Mendeley readers | | | |
|---|---|---|---|---|---|---|---|---|
| Sub-category* | aJIF | gJIF | aJIF 10+ | gJIF 10+ | aJIF | gJIF | aJIF 10+ | gJIF 10+ |
| Pharmacology, Toxicology and Pharmaceutics (misc.) | - | - | - | - | - | - | - | - |
| Drug Discovery | 0.852 | **0.872** | 0.871 | **0.883** | 0.794 | **0.798** | 0.900 | **0.910** |
| Pharmaceutical Science | 0.835 | **0.859** | 0.946 | 0.945 | 0.762 | **0.764** | 0.877 | **0.905** |
| Pharmacology | 0.817 | **0.832** | 0.911 | **0.920** | 0.775 | 0.775 | 0.884 | **0.903** |
| Toxicology | 0.830 | **0.833** | 0.897 | **0.911** | 0.816 | **0.823** | 0.859 | **0.889** |

*The highest out of the aJIF and gJIF is in bold, and the highest out of the Mendeley and Scopus gJIF is underlined.
+ The 10+ JIF variants exclude journals publishing less than 10 articles in a given year.
- The first category had no journals in the early years.

Table 5. Correlations over time between journal rankings based upon arithmetic and geometric JIFs for the Scopus category of Social Sciences. In each case the correlation is the average Spearman correlation between all ten pairs of consecutive years from 2004 to 2014.

| Sub-category* | Scopus citations | | | | Mendeley readers | | | |
|---|---|---|---|---|---|---|---|---|
| | aJIF | gJIF | aJIF 10+ | gJIF 10+ | aJIF | gJIF | aJIF 10+ | gJIF 10+ |
| Social Sciences (misc.) | **0.849** | 0.849 | 0.861 | **0.865** | 0.848 | <u>**0.852**</u> | 0.873 | **0.885** |
| Archeology | **0.839** | <u>0.836</u> | **0.903** | 0.901 | **0.824** | 0.817 | 0.894 | **0.905** |
| Development | 0.885 | <u>**0.887**</u> | 0.871 | **0.872** | 0.862 | <u>**0.876**</u> | 0.893 | **0.911** |
| Education | 0.749 | **0.759** | 0.869 | **0.874** | 0.762 | <u>**0.781**</u> | 0.880 | **0.892** |
| Geography, Planning and Development | 0.806 | <u>**0.806**</u> | 0.860 | **0.867** | 0.743 | <u>0.743</u> | 0.833 | **0.845** |
| Health (social science) | **0.845** | <u>0.845</u> | 0.879 | **0.888** | **0.832** | 0.830 | 0.855 | **0.865** |
| Human Factors and Ergonomics | 0.824 | **0.834** | 0.847 | **0.867** | 0.875 | <u>**0.886**</u> | 0.873 | **0.888** |
| Law | **0.787** | <u>0.783</u> | 0.868 | **0.876** | 0.727 | **0.740** | 0.885 | **0.890** |
| Library and Information Sciences | 0.845 | <u>**0.850**</u> | 0.881 | **0.889** | 0.821 | <u>**0.839**</u> | 0.871 | **0.895** |
| Linguistics and Language | **0.852** | 0.847 | 0.902 | **0.908** | 0.874 | <u>**0.874**</u> | 0.883 | **0.905** |
| Safety Research | 0.863 | <u>**0.864**</u> | 0.888 | **0.894** | 0.768 | **0.794** | 0.834 | **0.860** |
| Sociology and Political Science | 0.814 | **0.816** | 0.879 | **0.886** | 0.823 | <u>**0.825**</u> | 0.867 | **0.878** |
| Transportation | 0.924 | <u>**0.928**</u> | 0.912 | **0.919** | 0.902 | **0.913** | 0.895 | **0.915** |
| Anthropology | **0.804** | <u>0.799</u> | **0.878** | 0.873 | **0.784** | 0.772 | 0.871 | **0.881** |
| Communication | 0.807 | **0.815** | 0.823 | **0.831** | 0.816 | <u>**0.837**</u> | 0.865 | **0.889** |
| Cultural Studies | **0.754** | <u>0.752</u> | 0.845 | **0.849** | 0.710 | **0.711** | 0.812 | **0.837** |
| Demography | **0.828** | 0.807 | **0.881** | 0.866 | **0.837** | <u>0.830</u> | **0.778** | 0.773 |
| Gender Studies | 0.809 | **0.812** | 0.819 | **0.828** | 0.785 | 0.781 | 0.789 | **0.810** |
| Life-span and Life-course Studies | **0.810** | 0.796 | **0.880** | 0.870 | 0.850 | <u>**0.851**</u> | 0.874 | **0.885** |
| Political Science and International Relations | 0.807 | **0.812** | 0.836 | **0.843** | 0.791 | **0.799** | 0.842 | **0.857** |
| Public Administration | 0.802 | **0.809** | 0.832 | **0.847** | 0.823 | <u>**0.835**</u> | 0.848 | **0.867** |
| Urban Studies | **0.878** | <u>0.874</u> | 0.896 | **0.897** | 0.848 | <u>**0.852**</u> | 0.873 | **0.885** |

*The highest out of the aJIF and gJIF is in bold, and the highest out of the Mendeley and Scopus gJIF is underlined.
+ The 10+ JIF variants exclude journals publishing less than 10 articles in a given year.

A more detailed examination was made of one category, Insect Science, for which the (10+) gJIF was a substantial improvement on the (10+) aJIF, to see whether the cause could be tracked down to highly cited articles. As shown in Figure 1, for both the gJIF and the aJIF, the correlations are high for old journal volumes and low for the more recent ones (agreeing with: Stringer, Sales-Pardo, & Amaral, 2008). The gJIF is substantially more stable than the aJIF for three out of four recent years as well as for the period 2005-2006.

To investigate 2005-2006 in more detail, Figures 2 and 3 show aJIFs and gJIFs for individual journals for the period 2005-2007. From Figure 2, it is clear that the lower aJIF stability 2005-2006 is due to one journal, ranked second in 2005 (ranked 8[th] for gJIF) but mid-ranked in 2006. This journal, *Systematics and Biodiversity*, published one highly cited article in 2005, with 170 citations. The next most cited article in the journal received only 30 citations in 2005. *Systematics and Biodiversity* only published 10 articles in 2005 (its first year in Scopus in this category), allowing its one highly cited article to influence its aJIF considerably more than its gJIF.

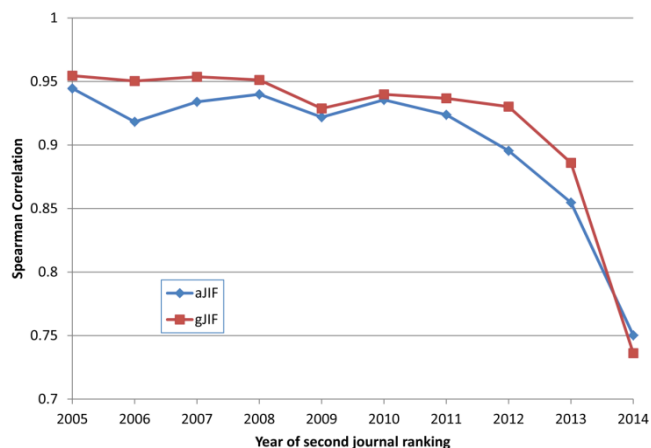

Figure 1. Spearman correlations between aJIF and gJIF rankings between consecutive years for journals with at least 10 articles in both years within the Insect Science category.

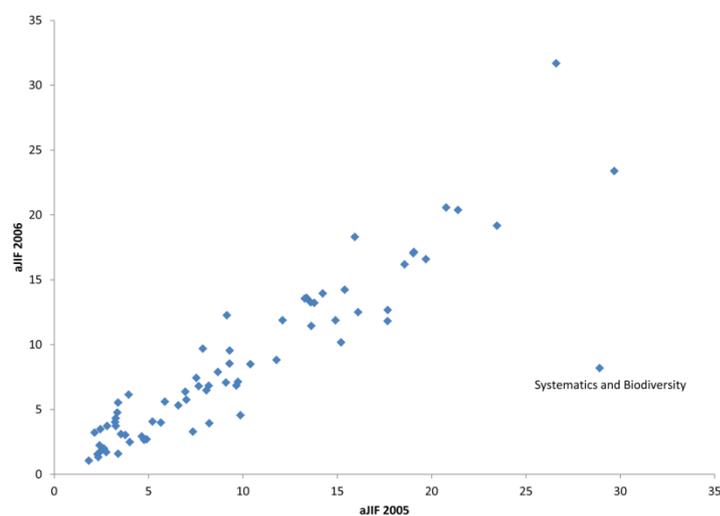

Figure 2. aJIF values for Insect Science journals with at least 10 articles in 2005 and 2006.

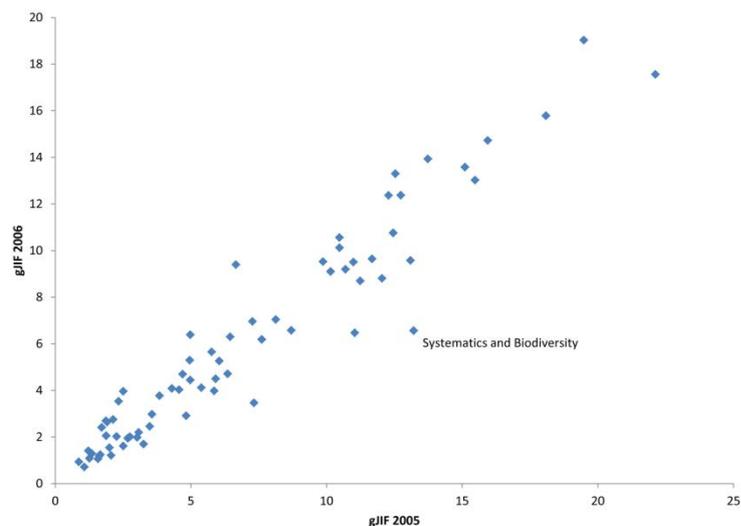

Figure 3. gJIF values for Insect Science journals with at least 10 articles in 2005 and 2006.

The other date during which the Insect Science aJIF differs most from the gJIF is 2011-12. From Figures 4 and 5, this discrepancy is again due to a single journal, *Fly*, and the cause is again a single highly cited article in the journal. This article, "A program for annotating and predicting the effects of single nucleotide polymorphisms, SnpEff: SNPs in the genome of Drosophila melanogaster strain w1118; iso-2; iso-3", had received 192 citations and the next most highly cited *Fly* article from 2012 had only 8 citations. This is a larger journal than *Systematics and Diversity*, with 52 articles in 2012, and so the relatively huge value of 192 did not disturb the gJIF much.

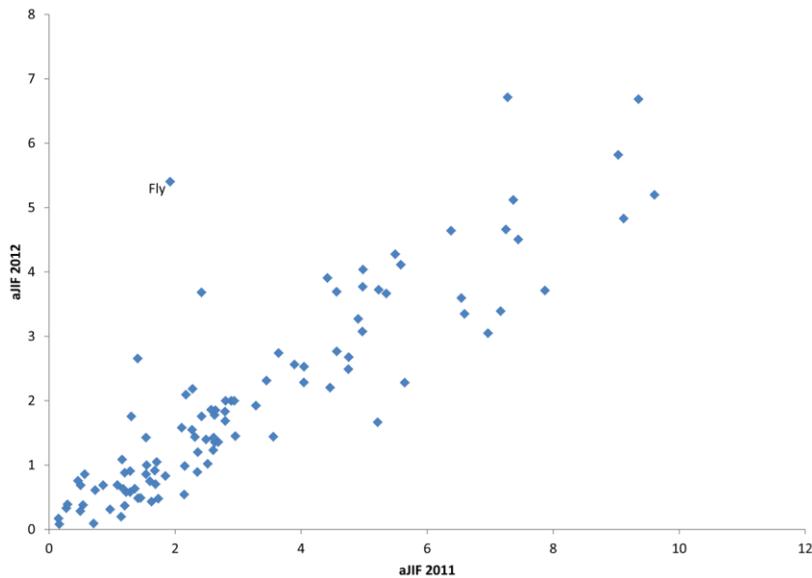
Figure 4. aJIF values for Insect Science journals with at least 10 articles in 2011 and 2012.

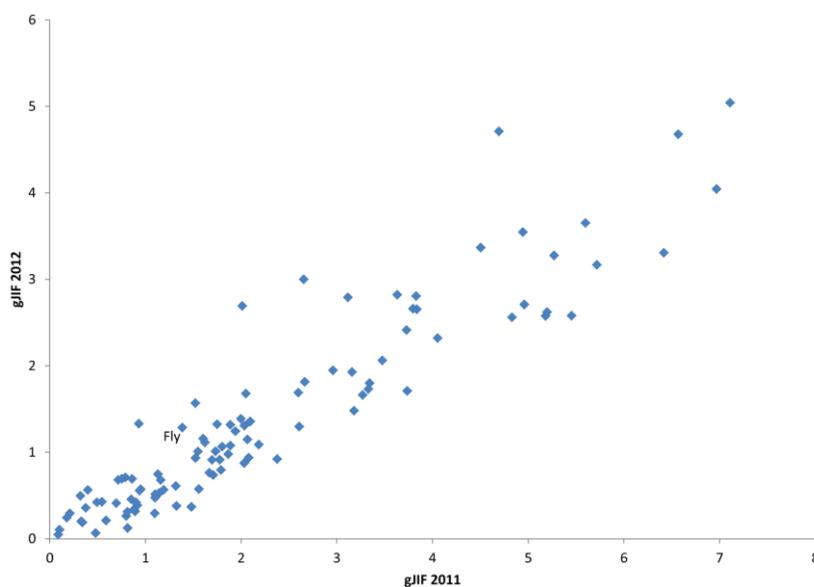
Figure 5. gJIF values for Insect Science journals with at least 10 articles in 2011 and 2012.

## Discussion and Conclusions

This study has several limitations. First, the coverage of disciplines is not comprehensive and, whilst it seems unlikely that there would be disciplines for which the gJIF is substantially less stable than the standard aJIF, there may be subject areas for which they are essentially the same, for example due to low overall citation counts. Second, the calculations tested are not the same as for the Thomson Reuters JIF, but are in a format that should accentuate the difference between the geometric and arithmetic variants of the JIF. Hence, the difference between the two is likely to be smaller than the numbers in the tables suggest. In addition, the assumption underlying the test is that the impact of a journal is relatively stable from year to year so that a high correlation between years indicates stability in the metric. Whilst this underlying journal stability has not been proven, intuitively it seems to be likely, in general. Nevertheless, there will be exceptions, such as journal direction changes, particularly with the appointment of a new editor, and with the publication of unusually high or low impact special issues. Nevertheless, these do not seem to be likely to give more stability to either variant of the JIF and so should not undermine the results.

The results show that the geometric JIF is superior to the standard JIF in terms of stability. It is presumably more resistant to changes due to individual highly cited articles temporarily increasing JIFs for a single year, unless there is a more subtle reason. Although the difference between the two JIFs is not large, since it is an improvement, it should be used in future in place of the standard JIF unless there are other reasons why the standard JIF is superior for a particular application. This conclusion applies to all broad fields since some fields had universally better results for the geometric JIF and none had universally better results for the arithmetic JIF. One disadvantage of the geometric JIF, however, is that it is less intuitive, especially for non-mathematical people, because the arithmetic mean is better known. This is also an advantage in the sense that it may discourage policy makers, research managers and scientists from considering it to be an obvious and intuitively correct measure of the impact of a journal. Additional analyses are needed, however, to assess whether the increase in stability of the gJIF has been made at the expense of other desirable, properties. For example, the standard JIF might more closely reflect peer judgements of journal quality if scientists valued the ability of a journal to occasionally produce highly cited articles.

The relatively small differences in stability between the aJIF and gJIF are perhaps surprising, given the prominence of the stability issue in DORA. This is especially true because the aJIF variant was calculated in a way to enhance its instability by considering only citations to articles from a single year. The Insect Science example confirms that the gJIF can ameliorate the effect of individual highly cited articles on a journal's ranking. Nevertheless, the small stability differences between the aJIF and gJIF suggest that individual highly cited articles in journals with low JIFs are rare. Hence, whilst there is at least one well-known example of this, and the impact on journal rankings can be stark (Foo, 2013), it should be acknowledged that, journal rankings tend to be relatively stable and this phenomenon is unusual rather than a normal part of science.

There are variations by discipline in the extent to which the gJIF is more stable than the JIF variant calculated for the table. Nevertheless, the variations do not seem to be large or systematic enough to suggest the presence of underlying disciplinary differences, except perhaps differences due to the amount of data available for the calculations – assuming that more data would tend to give more consistent results. There are also disciplinary differences in the extent to which Mendeley readership-based journal rankings are more stable (or less stable) than Scopus citation-based journal rankings. These differences seem likely to be due to differing relative amounts of citation and readership data, on the basis that the results are likely to be more stable for data based on larger counts.

The same conclusion is also valid for Mendeley-based JIFs: the geometric mean variant is more stable, and hence likely to be a better indicator of impact, than the standard arithmetic mean variant, unless standard JIFs are superior in terms of other properties, such as correlation with peer judgements. In addition, the magnitudes of the Mendeley-based correlations are similar to those for the Scopus citation-based JIFs, which is evidence of their stability and, hence is evidence that Mendeley reader counts consistently indicate the same type of impact, which is presumably academic impact with a bias towards more junior scholars.

# References

ABS (2010). Academic journal quality guide version 4. http://www.associationofbusinessschools.org/sites/default/files/Combined%20Journal%20Guide.pdf

Adler, N. J., & Harzing, A. W. (2009). When knowledge wins: Transcending the sense and nonsense of academic rankings. Academy of Management Learning & Education, 8(1), 72-95.

Adie, E., & Roe, W. (2013). Altmetric: enriching scholarly content with article-level discussion and metrics. Learned Publishing, 26(1), 11-17.

Al-Awqati, Q. (2007). Impact factors and prestige. Kidney International, 71(3), 183-185.

Archambault, É., & Larivière, V. (2009). History of the journal impact factor: Contingencies and consequences. Scientometrics, 79(3), 635-649.

Bartol, T., Budimir, G., Dekleva-Smrekar, D., Pusnik, M., & Juznic, P. (2014). Assessment of research fields in Scopus and Web of Science in the view of national research evaluation in Slovenia. Scientometrics, 98(2), 1491-1504.

Baum, J. A. (2011). Free-riding on power laws: Questioning the validity of the impact factor as a measure of research quality in organization studies. Organization, 18(4), 449-466.

Bergstrom, C. T., West, J. D., & Wiseman, M. A. (2008). The Eigenfactor™ metrics. The Journal of Neuroscience, 28(45), 11433-11434.

Brown, H. (2007). How impact factors changed medical publishing—and science. Bmj, 334(7593), 561-564.

CORE (2014). CORE Conference rankings process 2014. http://www.core.edu.au/documents/CORE2014RankingsInstructions.pdf

Donohue, J. M., & Fox, J. B. (2000). A multi-method evaluation of journals in the decision and management sciences by US academics. Omega, 28(1), 17-36.

DORA (2012). San Francisco Declaration on Research Assessment. http://am.ascb.org/dora/

Foo, J. Y. A. (2013). Implications of a single highly cited article on a journal and its citation indexes: A tale of two journals. Accountability in Research, 20(2), 93-106.

Franceschet, M., & Costantini, A. (2011). The first Italian research assessment exercise: A bibliometric perspective. Journal of Informetrics, 5(2), 275-291.

Greenwood, D. C. (2007). Reliability of journal impact factor rankings. BMC Medical Research Methodology, 7(1), 48. doi:10.1186/1471-2288-7-48

Haddow, G., & Genoni, P. (2010). Citation analysis and peer ranking of Australian social science journals. Scientometrics, 85(2), 471-487.

Haustein, S., & Siebenlist, T. (2011). Applying social bookmarking data to evaluate journal usage. Journal of Informetrics, 5(3), 446-457.

Kalaitzidakis, P., Mamuneas, T. P., & Stengos, T. (2003). Rankings of academic journals and institutions in economics. Journal of the European Economic Association, 1(6), 1346-1366.

Kalaitzidakis, P., Mamuneas, T. P., & Stengos, T. (2011). An updated ranking of academic journals in economics. Canadian Journal of Economics, 44(4), 1525-1538.

Larivière, V., & Gingras, Y. (2010). The impact factor's Matthew effect: A natural experiment in bibliometrics. Journal of the American Society for Information Science and Technology, 61(2), 424-427.

Li, X., Thelwall, M., & Giustini, D. (2012). Validating online reference managers for scholarly impact measurement, Scientometrics, 91(2), 461-471.

13Lozano, G. A., Larivière, V., & Gingras, Y. (2012). The weakening relationship between the impact factor and papers' citations in the digital age. Journal of the American Society for Information Science and Technology, 63(11), 2140-2145.

Lundberg, J. (2007). Lifting the crown—citation z-score. Journal of Informetrics, 1(2), 145-154.

MacRoberts, M. H., & MacRoberts, B. R. (1989). Problems of citation analysis: A critical review. Journal of the American Society for Information Science, 40(5), 342-349.

Maflahi, N. & Thelwall, M. (in press). When are readers as good as citers for bibliometrics? Scopus vs. Mendeley for LIS journals. Journal of the Association for Information Science and Technology.

Moed, H. F., & Visser, M. S. (2008). Appraisal of citation data sources. Centre for Science and Technology Studies, Leiden University. http://www.hefce.ac.uk/media/hefce/content/pubs/indirreports/2008/missing/Appraisal%20of%20Citation%20Data%20Sources.pdf.

Mohammadi, E., Thelwall, M., Haustein, S., & Larivière, V. (in press). Who reads research articles? An altmetrics analysis of Mendeley user categories. Journal of the Association for Information Science and Technology.

Mohammadi, E., Thelwall, M. & Kousha, K. (in press). Can Mendeley bookmarks reflect readership? A survey of user motivations. Journal of the Association for Information Science and Technology.

Moussa, S., & Touzani, M. (2010). Ranking marketing journals using the Google Scholar-based hg-index. Journal of Informetrics, 4(1), 107-117.

Nisonger, T. E., & Davis, C. H. (2005). The perception of library and information science journals by LIS education deans and ARL library directors: A replication of the Kohl–Davis study. College & Research Libraries, 66(4), 341-377.

Priem, J., Taraborelli, D., Groth, P., & Neylon, C. (2010). Altmetrics: A manifesto. http://altmetrics.org/

Rainer Jr, R. K., & Miller, M. D. (2005). Examining differences across journal rankings. Communications of the ACM, 48(2), 91-94.

Robinson, M. D., & Smyth, G. K. (2008). Small-sample estimation of negative binomial dispersion, with applications to SAGE data. Biostatistics, 9(2), 321-332.

Seglen, P. O. (1992). The skewness of science. Journal of the American Society for Information Science, 43(9), 628-638.

Seglen, P. O. (1994). Causal relationship between article citedness and journal impact. Journal of the American Society for Information Science, 45(1), 1-11.

Seglen, P. O. (1997). Why the impact factor of journals should not be used for evaluating research. BMJ, 314(7079), 497.

Serenko, A., & Dohan, M. (2011). Comparing the expert survey and citation impact journal ranking methods: Example from the field of Artificial Intelligence. Journal of Informetrics, 5(4), 629-648.

Shema, H., Bar-Ilan, J., & Thelwall, M. (2014). Do blog citations correlate with a higher number of future citations? Research blogs as a potential source for alternative metrics. Journal of the Association for Information Science and Technology, 65(5), 1018-1027.

Shepherd, P. (2007). The feasibility of developing and implementing journal usage factors: a research project sponsored by UKSG. Serials: The Journal for the Serials Community, 20(2), 117-123.

Steward, M. D., & Lewis, B. R. (2010). A comprehensive analysis of marketing journal rankings. Journal of Marketing Education, 32(1), 75-92.

Stringer, M. J., Sales-Pardo, M., & Amaral, L. A. N. (2008). Effectiveness of journal ranking schemes as a tool for locating information. PloS ONE, 3(2), e1683.

Thelwall, M., Haustein, S., Larivière, V., & Sugimoto, C. R. (2013). Do altmetrics work? Twitter and ten other social web services. PloS ONE, 8(5), e64841.

Thelwall, M. & Wilson, P. (2014). Distributions for cited articles from individual subjects and years. Journal of Informetrics, 8(4), 824-839.

Thelwall, M. & Wilson, P. (in press). Mendeley readership Altmetrics for medical articles: An analysis of 45 fields, Journal of the Association for Information Science and Technology.

Thelwall, M. (2012). Journal impact evaluation: a webometric perspective. Scientometrics, 92(2), 429-441.

Thomson Reuters (2014). The Thomson Reuters Impact Factor. http://wokinfo.com/essays/impact-factor/

Torres-Salinas, D., Lopez-Cózar, E. D., & Jiménez-Contreras, E. (2009). Ranking of departments and researchers within a university using two different databases: Web of Science versus Scopus. Scientometrics, 80(3), 761-774.

UNSW (2015). Excellence in Research for Australia (ERA) outlet ranking. https://research.unsw.edu.au/excellence-research-australia-era-outlet-ranking

Vanclay, J. K. (2012). Impact factor: outdated artefact or stepping-stone to journal certification? Scientometrics, 92(2), 211-238.

Zahedi, Z., Costas, R., & Wouters, P. (2014). How well developed are altmetrics? A cross-disciplinary analysis of the presence of 'alternative metrics' in scientific publications. Scientometrics, 101(2), 1491-1513.

Zahedi, Z., Haustein, S. & Bowman, T (2014). Exploring data quality and retrieval strategies for Mendeley reader counts. Presentation at SIGMET Metrics 2014 workshop, 5 November 2014. http://www.slideshare.net/StefanieHaustein/sigmetworkshop-asist2014

Zitt, M. (2012). The journal impact factor: Angel, devil, or scapegoat? A comment on JK Vanclay's article 2011. Scientometrics, 92(2), 485-503.